\documentstyle[12pt,a4wide,epsfig]{article}
\begin{document}
\parindent 1.3cm
\thispagestyle{empty}   
\vspace*{-3cm}
\noindent
\vspace{2.5cm}
\def\ds{\displaystyle}
\begin{flushright}
NORDITA 97/10 N,P\\
UAB-FT-97/411\\
hep-ph/9702302
\end{flushright}

\vspace{1cm}

\begin{center}
\begin{bf}
CHIRAL--LOOP AND VECTOR--MESON CONTRIBUTIONS\\
TO $\eta \to \pi \pi \gamma \gamma $ DECAYS
\end{bf}
\vspace{2.5cm}\\
LL. AMETLLER$^a$, J. BIJNENS$^b$,  A. BRAMON$^c$ and P. TALAVERA$^b$
\vspace{0.8cm}\\

$^a$ Deptartament de F{\'\i}sica i Enginyeria Nuclear, \\
Universitat Polit\`ecnica de Catalunya, 08034 Barcelona, Spain\\
~\vspace{0.1cm}\\
$^b$ NORDITA, Blegdamsvej 17, DK 2100 Copenhagen, Denmark\\
~\vspace{0.1cm}\\
$^c$ Departament de  F{\'\i}sica, Universitat Aut\`onoma de Barcelona,\\ 
08193 Bellaterra (Barcelona), Spain\\
~\vspace{0.1cm}\\

\vspace{2.4cm}

{\bf ABSTRACT}

\end{center}

The process $\eta \to \pi^0 \pi^0\gamma \gamma$ is discussed 
in Chiral Perturbation Theory (ChPT).
Special attention is devoted to one-loop 
corrections, $\eta$-$\eta'$ mixing effects and vector-meson 
dominance of ChPT counter-terms. The less interesting 
$\eta \to \pi^+ \pi^-\gamma \gamma$ transition 
is briefly discussed too.

\vspace{3cm}

\newpage
\setcounter{page}{1}
The rare decay modes  $\eta \to \pi^0 \pi^0\gamma \gamma$ and
$\eta \to \pi^+ \pi^-\gamma \gamma$ have been recently discussed  
by several authors
in the context of Chiral Perturbation Theory (ChPT)
\cite{GL}. A lowest order analysis, i.e. at tree-level in ChPT, has been
performed for both decays by Kn\"ochlein, Scherer and Drechsel
\cite{KNOCH}. More recently, Bellucci and Isidori \cite{BI} have extended
the analysis for the neutral mode to one-loop in ChPT.
When that reference appeared we had finished our analytical work. Preliminary
results of this work were presented in \cite{upsala}.
The potential
interest of these rare decays is in view of the large number of $\eta$'s 
to be produced in a near future by several facilities. 
Moreover, the necessity of computing the 
large one-loop corrections can be traced to the 
closely related analysis for $\gamma\gamma\to 3\pi$ \cite{TABBC},
as it has been anticipated \cite{BI,upsala} and
fully confirmed in this work and other recent results\cite{BI}.

We would like to add that our previous
computation and discussion of the $\eta\to\pi^0\gamma\gamma$ 
amplitude in ChPT \cite{ABBC} is also particularly 
illustrating when computing $\eta\to\pi^0\pi^0\gamma\gamma$. 
Indeed, the observation in \cite{ABBC} that the main contribution 
to $A(\eta\to\pi^0\gamma\gamma)$ comes from the vector
meson dominated (VMD) counter-terms of the ChPT Lagrangian, 
strongly suggests the possibility of similar important contributions to
$A(\eta\to\pi^0\pi^0\gamma\gamma)$. 
The main purpose of the present note is to
reconsider the $\eta\to\pi^0\pi^0\gamma\gamma$ decay including this
contribution and refining some of the findings in refs.\cite{KNOCH,BI}.
The $\eta \to \pi^+ \pi^-\gamma \gamma$ amplitude will be briefly 
discussed too.

Our notation follows closely that in refs.\cite{TABBC,ABBC,HANS,DAPHNE} 
to which we refer
for details. In this notation and for later use, we quote 
a few relevant amplitudes involving vertices from the 
lowest-order piece of the ChPT lagrangian, ${\cal L}^{(2)}$,
\begin{eqnarray}
& A(\eta_8\to \pi^0\pi^0\pi^0) = - B (m_d - m_u)/\sqrt{3} f^2 = -2 \Delta m_K^2
/\sqrt{3} f^2  \nonumber \\
& A(\pi^+ (p_+) \pi^-(p_-) \to \pi^0(p_1) \pi^0(p_2))= 2 (s_{\pi\pi}-
m_\pi^2)/f^2,
\end{eqnarray}
where $s_{\pi\pi}\equiv (p_++p_-)^2=(p_1+p_2)^2 = p_{12}^2$, 
$f_\pi\simeq f=132$~MeV, and $\Delta m_K^2\simeq 6.2\times 10^{-3}$~GeV$^{-2}$ is the
non-photonic contribution to the kaon mass difference (see \cite{ABBC}).
Other useful amplitudes involving vertices of the anomalous sector
of the ChPT lagrangian, ${\cal L}^{(4)}_{WZ}$, are, e.g.,
\begin{eqnarray}
\label{ANOM}
& A(\pi^0\to \gamma(k_1)\gamma(k_2))= \ds \sqrt{3} A(\eta_8 \to \gamma\gamma)=
{-\sqrt{2} e^2\over 4 \pi^2 f} \epsilon_{\mu\nu\alpha\beta}
\epsilon_1^\mu k_1^\nu \epsilon_2^\alpha k_2^\beta \nonumber\\
& A(\eta_8 (p_3) \to \pi^+(p_+) \pi^-(p_-) \gamma)= 
\ds {e\over \sqrt{6} \pi^2 f^3}
\epsilon_{\mu\nu\alpha\beta} \epsilon^\mu p_3^\nu p_+^\alpha p_-^\beta.
\end{eqnarray}

The use of VMD to estimate counter-terms of the anomalous sector of 
the ChPT lagrangian (see ref.\cite{HANS})
requires the introduction of effective lagrangians such as
$${\cal L}_{VVP}= G/\sqrt{2} \epsilon_{\mu\nu\alpha\beta} 
tr (\partial^\mu V^\nu \partial^\alpha V^\beta P)$$ 
and
$${\cal L}_{V\gamma}= -2 e g f^2 A^\mu tr (Q V_\mu)= 
-e M_{\rho,\omega}^2 /\sqrt{2} g (\rho_\mu^0 A^\mu+
{1\over 3}\omega_\mu A^\mu - {\sqrt{2}\over 3} \phi_\mu A^\mu)$$ 
accounting for the $VVP$ and $V\gamma$
transitions with $M_\rho^2\simeq M_\omega^2 = 2 g^2 f^2$ 
and $g\simeq 4.15$ \cite{HANS,DAPHNE}. From 
these lagrangians one obtains a series of SU(3)-related 
amplitudes such as
\begin{eqnarray}
\label{vpg}
&A(\omega\to\rho^0\pi^0)=\sqrt{3} A(\omega\to \omega\eta_8)= G
\epsilon_{\mu\nu\alpha\beta} \epsilon_i^\mu p_i^\nu \epsilon_f^\alpha
p_f^\beta \nonumber\\
& A(\omega\to\pi^0\gamma)= 3 A(\rho\to\pi\gamma)= \ds {e G \over \sqrt{2} g}
\epsilon_{\mu\nu\alpha\beta}\epsilon_i^\mu p_i^\nu \epsilon_f^\alpha 
k^\beta, 
\end{eqnarray}
which allows one to recover the whole standard VMD phenomenology 
and, in particular, the previous amplitudes
(\ref{ANOM}) for $\pi^0, \eta_8 \to \gamma\gamma$ if 
$G=3\sqrt{2} g^2/4\pi^2 f$.

The treatment in ChPT of the physical $\eta$-particle, 
resulting from the mixing  between the octet and singlet 
states $\eta_8$ and $\eta_1$, is not a
trivial subject and deserves special attention, as recently emphasized 
by Leutwyler and others\cite{PERIS,LEUT}. Following these references and 
previous phenomenological work in refs \cite{ABBC,HANS},
we are going to distinguish between radiative vertices
related to the anomalous part of the lagrangian, ${\cal L}_{WZ}^{(4)}$,
and the non-anomalous ones coming from  ${\cal L}^{(2)}$ and
${\cal L}^{(4)}$. In the latter case,
only the $SU(3)$ octet of Goldstone bosons are considered to appear in
${\cal L}^{(2)}$ and the whole effect of the octet-singlet mixing is assumed to
proceed through the $L_7^r$ counter-term of the next order piece of the
lagrangian, ${\cal L}^{(4)}$ \cite{GL,PERIS,LEUT}. As it is suggested in \cite{LEUT},
however, the situation can be different for the radiative vertices contained
in ${\cal L}_{WZ}^{(4)}$. Indeed, a satisfactory description of $\eta\to
\gamma\gamma$ and $\eta\to \pi^+ \pi^-\gamma$ can only be achieved
\cite{HANS,BBC} by introducing the phenomenological $\eta$~-~$\eta'$ 
mixing angle, $\theta\simeq -19.5^\circ$ \cite{GILMAN}, 
and extending from the $SU(3)$-octet to
the $U(3)$-nonet the fields appearing in ${\cal L}_{WZ}^{(4)}$. In this
extended nonet-symmetric context one then has
$A(\eta_1\to\gamma\gamma)=2\sqrt{2}A(\eta_8 \to\gamma\gamma)$,
$A(\eta_1\to \pi^+\pi^-\gamma)= \sqrt{2} A(\eta_8 \to \pi^+\pi^-\gamma)$,
$A(\rho\to \eta_1\gamma)= \sqrt{2} A(\rho\to \eta_8\gamma)$, and other
similarly simple relations for anomalous vertices.

Once we have fixed all the above couplings, the computation of the $\eta\to
\pi^0\pi^0\gamma\gamma$ amplitude in ChPT is a straightforward
task. For convenience we will consider four separate contributions to the
amplitude  
\begin{eqnarray}
\label{4contrib}
A(\eta\to \pi^0\pi^0\gamma\gamma) &\equiv & A(\eta(p_3)\to
 \pi^0(p_1)\pi^0(p_2)\gamma(k_1) \gamma(k_2)) \nonumber \\
 & = &A(\eta\to \pi^0\pi^0\gamma\gamma)_{\pi^0-pole}+
 A(\eta\to \pi^0\pi^0\gamma\gamma)_{\eta-tail}\nonumber \\
 & + &
 A(\eta\to \pi^0\pi^0\gamma\gamma)_{1PI}+
  A(\eta\to \pi^0\pi^0\gamma\gamma)_{VMD}.
\end{eqnarray}
The first two correspond to one-particle reducible 
diagrams containing the $\pi^0$ or $\eta$ propagator 
and will be computed up to one-loop, i. e., at $O(p^4)$ and $O(p^6)$
in ChPT, as in ref.\cite{BI}. The third term corresponds to
one-particle irreducible (1PI) diagrams and completes the one-loop
calculation, $O(p^6)$ in ChPT. The final term contains the VMD
contributions to the low energy constants or counter-terms of the ChPT
Lagrangian at order $p^8$  and higher, but their effects are not necessarily
negligible, as the analysis of $\eta\to \pi^0 \gamma\gamma$ in \cite{ABBC}
indicates.

Most of the  $\eta\to \pi^0\pi^0\gamma \gamma$
decay events are expected to proceed through the $\eta\to \pi^0
\pi^0  \pi^0 \to \pi^0  \pi^0 \gamma \gamma$ decay chain, having to do with
the well-studied, isospin-violating 
$\eta\to 3\pi^0$ decay amplitude \cite{GL, LEUT}, rather
than being genuine $\eta\to \pi^0\pi^0 \gamma\gamma$ decays. A  convenient
way to write the ChPT amplitude for this kind of background is
\begin{eqnarray}
\label{pi0pole}
 A( \eta\to\pi^0\pi^0\gamma\gamma)_{\pi^0-pole}= && \ds {-e^2 \over
\sqrt{6} \pi^2 f_\pi^3}~ {\Delta m_K^2
\over s - m_\pi^2 + i m_\pi \Gamma_\pi}~
\epsilon_{\mu\nu\alpha\beta}
\epsilon_1^\mu k_1^\nu \epsilon_2^\alpha k_2^\beta \nonumber\\ 
&& \ds \times (1 + U + V + W) 
(1 + {1\over 3} {m_\pi^2\over m_\eta^2 -m_\pi^2}
-{1\over 3}{m_\pi^2 \over m_\eta^2-s}),
\end{eqnarray}
where $s\equiv s_{\gamma\gamma}=(k_1+k_2)^2$ and the final factor 
has been included to account for the off-shellness 
of the $\pi^0$ in one of the three possible
$\pi^0\to \gamma\gamma$ transitions, but can be safely neglected.
Taking $U=V=W=0$ and $f_\eta=f_\pi=f$ in the above amplitude corresponds to
the $O(p^4)$  tree-level result. 
Sizeable loop corrections and $\eta$-$\eta'$ effects
are included following refs. \cite{GL, LEUT}. For the former we 
simplify our analysis (see ref. \cite{PETRUS}) taking 
$U+V=0.39 -0.03 + 0.18 i$, as corresponds to the center 
of the $\eta\to 3\pi^0$ Dalitz plot \cite{GL}. 
The $\eta$-$\eta'$ mixing effects are
parametrized taking $1+W\simeq 1 + {2\over 3} \Delta_{GMO} \simeq 1.15$
in agreement with the value given in \cite{GL,LEUT}, but somewhat below
$1+W\simeq \sqrt{2}$ coming from 
nonet-symmetry arguments with $\theta\simeq -19.5^\circ$.
With these values and $\Delta m_K^2
\simeq 6.2 \times 10^{-3}$~GeV$^2$ one obtains $\Gamma(\eta\to 3\pi^0)=
315$~eV, reasonably close to the experimental value 
$\Gamma(\eta\to~3\pi^0)_{exp}=379 \pm 36$~eV \cite{PDG}. 
Using eq.(\ref{pi0pole})
one also obtains the dashed curve plotted in Fig. 1, clearly showing the
$\pi^0$-pole in the $\gamma\gamma$ mass spectrum. Both this result and the
above discussion are essentially equivalent to those presented
in ref.\cite{BI}, except for a possible sign (see below) of no relevance
here and the value for $(1+U+V+W) \simeq \rho =2$ which in \cite{BI} is 
taken from the experimental $\eta\to 3\pi^0$ decay width.

Another independent contribution to $\eta\to \pi^0\pi^0\gamma\gamma$
proceeds through the isospin conserving decay chain $\eta\to\pi^0\pi^0\eta
\to\pi^0\pi^0\gamma\gamma$. The corresponding ChPT amplitude at one-loop can
be written as
\begin{eqnarray}
\label{etatail}
A(\eta\to \pi^0\pi^0\gamma\gamma)_{\eta-tail} & = \ds {e^2  \over 12
\sqrt{6} \pi^2 f_\pi^2 f_\eta}~ {m_\pi^2 \over s - m_\eta^2}~
(1 + C_{loops} )~ (\cos\theta - 2\sqrt{2}
\sin\theta) \nonumber\\
& \ds \times \epsilon_{\mu\nu\alpha\beta}  
\epsilon_1^\mu k_1^\nu \epsilon_2^\alpha k_2^\beta + 
{k_1\choose \epsilon_1} \leftrightarrow {k_2\choose \epsilon_2}
+(p_1\leftrightarrow p_2)
+ \cdots,
\end{eqnarray}
where the angular factor $\cos\theta -2 \sqrt{2}\sin\theta \simeq 4
\sqrt{2}/3$ takes into account both the $\eta_8$ and $\eta_1\to\gamma\gamma$
radiative decays thus justifying the use of the physical $\eta$ mass in the
propagator. The dots stand for negligible contributions involving
an $\eta'$ propagator. As
previously discussed, the treatment of $\eta$-$\eta'$ mixing for the initial
$\eta$ ---coupled to $\pi^0 \pi^0 \eta$ through the mass terms in ${\cal
L}^{(2)}$--- is different and leads to the factor $(1 + C_{loops} )$ in the
above amplitude where, apart from the tree-level contribution, the loop
effects are included in $C_{loops}$. They are explicitly given in the 
Appendix and contain
$\eta$~-$\eta'$ mixing effects in the $L_7^r$ counter-term \cite{LEUT}. A more
sophisticated analysis of this contribution to $\eta\to
\pi^0\pi^0\gamma\gamma$ seems unnecessary since its numerical effects, as
also observed in \cite{BI}, are found to be rather small.

A more genuine contribution to $\eta\to \pi^0\pi^0\gamma\gamma$ involves
pion loops (as well as numerically less important kaon, $\eta$ and 
$\pi\eta$ loops) with an $\eta\to \pi^+ \pi^-\gamma (\gamma)$ 
anomalous vertex from ${\cal L}_{WZ}^{(4)}$ followed by
$\pi^+\pi^- (\gamma) \to \pi^0 \pi^0 \gamma$ rescattering. As expected
this is an important correction. The reason is that, contrary
to the tree-level amplitudes,  
this $O(p^6)$ contribution does not vanish in the chiral limit.
Indeed, restricting for the moment to the octet 
piece of the physical $\eta$, we find
\vfill
\newpage
\begin{eqnarray}
\label{1pi}
 A(\eta_8&\to&\pi^0\pi^0\gamma\gamma)^{\pi-loops}_{1PI}=
 \ds {- 4 e^2 (m_\pi^2 - p_{12}^2 ) \over \sqrt{6} \pi^2 f_\pi^2 f_\eta}
{1\over 16  \pi^2 f^2}  \\
&& 
\times R(p_{12}^2, -k\cdot p_{12},m_\pi^2) \epsilon_{\mu\nu\alpha\beta}
\bigl(-\epsilon_1^\mu + {\epsilon_1\cdot p_{12} \over k_1\cdot p_{12}}
k_1^\mu \bigr) k_2^\nu p_3^\alpha \epsilon_2^\beta + 
\Bigl[{k_1\choose \epsilon_1} \leftrightarrow {k_2\choose \epsilon_2}
\Bigr],\nonumber
\end{eqnarray}
where the function $R$ is defined in the Appendix. This result 
---which is pivotal in the whole discussion--- fully agrees with
the expression obtained by Bellucci and Isidori (BI) \cite{BI}. 

However, when eq.(\ref{1pi}) is 
combined with the previous $\pi^0$--pole and $\eta$--tail amplitudes, 
eqs.(\ref{pi0pole},\ref{etatail}), we
disagree in the interference pattern. While in \cite{BI}, there is 
constructive interference in the interval $0.08 \le z \equiv
m_{\gamma\gamma}^2/m_\eta^2 < 0.25$, we obtain a destructive one. 
We believe that the origin of this discrepancy is a different, 
relative sign between our $\pi^0$--pole contribution (\ref{pi0pole}) 
and the corresponding one in ref.\cite{BI}, once the two notations 
have been unified. 
In Fig. 1 we plot (solid line) the di-photon mass spectrum obtained adding
our eqs.(\ref{pi0pole}) and (\ref{1pi}). To approximately 
reproduce the results by BI \cite{BI}, 
we simply have to reverse the sign of our eq.(\ref{pi0pole}).
The resulting curve is also plotted (upper dotted
line) in Fig. 1.

Up to this point, in this 1PI contribution
there has been no discussion 
on the effects of $\eta$-$\eta'$ mixing, which, as previously 
mentioned, are required to correctly describe
the $\eta\to \pi^+\pi^-\gamma (\gamma)$ vertices \cite{BBC} appearing in 
the 1PI one-loop diagrams. 
Restricting to the dominant pion loops, the $\eta$-$\eta'$
mixing effects translate simply into the following enhancement in 
the amplitude for the physical $\eta$
\begin{eqnarray}
\label{1pietaphys}
A(\eta\to \pi^0\pi^0\gamma\gamma)_{1PI}^{\pi-loops} &=&
(\cos\theta - \sqrt{2} \sin\theta) \times A(\eta_8 \to
\pi^0\pi^0\gamma\gamma)_{1PI}^{\pi-loops}\nonumber \\
& \simeq & \sqrt{2} A(\eta_8 \to
\pi^0\pi^0\gamma\gamma)_{1PI}^{\pi-loops}.
\end{eqnarray} For completeness, we have also computed kaon and $\pi\eta$
loop effects confirming that they are small compared to the 
dominant one from pion loops, as also found in \cite{BI} for the
kaon loops ($\pi\eta$ loops were neglected there).
The complete expressions for the $\eta$-tail and 1PI 
amplitudes at  $O(p^6)$ are given in the Appendix. 
Notice the presence of the mixing angle $\theta$ and the low-energy
constant $L_7^r$ in our result for the $\eta$--tail amplitude. 
This is not double-counting the mixing effects, but it is due to
our treatment for the $\eta$--$\eta'$ mixing, using explicitly the angle
$\theta$ in the radiative  transitions and $L_7^r$ in the non-anomalous
vertices involving four pseudoscalars. The rest of $L_i^r$ counter-terms
are needed to obtain a finite result, which turns out to be rather
insensitive to their actual values.
We show in Fig. 2 the di-photon spectrum corresponding to our $O(p^6)$
result, taking $f_\pi\simeq f$ and $f_\eta\simeq 1.3 f_\pi$  
(upper dotted line). It turns out that these 
complete one-loop amplitude dominates over the
$\pi^0$--pole background alone for $z < m_\pi^2/m_\eta^2$ (at the 
left-hand side of the pion-pole) and also for $z > 0.17$. (This result 
contrasts again with the one obtained by BI.) 

At this point, one can wonder about the importance of next-next-to leading 
corrections. This corresponds to $O(p^8)$ in the chiral counting and its
complete computation is out of the scope of the present work. It would involve
two-loop diagrams with one vertex from ${\cal L}^{(4)}_{WZ}$ and the rest
from ${\cal L}^{(2)}$; one-loop diagrams with either one vertex from  ${\cal
L}^{(6)}_{WZ}$ and the rest from ${\cal L}^{(2)}$ or with the presence of
vertices from ${\cal L}^{(4)}_{WZ}$ and ${\cal L}^{(4)}$;
and also $O(p^8)$ tree-level diagrams. The latter can be estimated through
VMD, which can also be viewed as a full all-order estimate
of the counter-terms, as discussed in \cite{ABBC} for the process
$\eta\to\pi^0\gamma\gamma$. Computing the VMD diagrams with two vector
meson propagators and using amplitudes like those in 
eq.(\ref{vpg}), one obtains
\begin{eqnarray}
\label{VMD}
A(\eta\to\pi^0\pi^0\gamma\gamma)_{VMD}&=&
\ds {e^2 G^3\over  \sqrt{6}g^2} (1+ {1\over 9})
\epsilon_{\rho\alpha\beta\gamma} \epsilon_{\lambda \sigma}^{\phantom{\lambda
\sigma}\rho\omega}
\epsilon_{\mu\nu\tau\omega} 
\epsilon_2^\alpha k_2^\beta (p_1+k_1)^\lambda \epsilon_1^\mu k_1^\nu
p_1^\tau \nonumber \\ &&\ds \Bigl({1\over m_V^2 -(p_3-k_2)^2} 
~ {1\over m_V^2 -(p_1+k_1)^2}
p_3^\gamma p_2^\sigma  \nonumber \\ 
&&\ds+ {1\over m_V^2 -(p_1+k_1)^2} ~ {1\over m_V^2
-(p_2+k_2)^2} p_2^\gamma p_3^\sigma \Bigr)
 \nonumber \\
&&\ds + 
\Bigl[{k_1\choose \epsilon_1} \leftrightarrow {k_2\choose \epsilon_2}
\Bigr] + (p_1 \leftrightarrow p_2),
\end{eqnarray}
In Fig. 2 we show (lower dotted curve) 
the di-photon invariant mass spectrum corresponding to
this VMD amplitude alone. Integrating over the whole spectrum leads to
$\Gamma_{VMD}(\eta\to \pi^0\pi^0\gamma\gamma) \simeq 3.2 \times
10^{-6}$~eV, not far from old VMD estimates \cite{BB}. 
Our final result, containing the four contributions listed in 
eq.(\ref{4contrib}), is also plotted (solid line) in Fig. 2 and can 
be compared with our previous full amplitude at order $p^6$ 
(upper dotted line). 
The higher order VMD contribution decreases (increases) 
the order--$p^6$ amplitude in the large (narrow) region of the 
$\gamma\gamma$--spectrum above (below) the $\pi^0$-pole. We predict a
rather small $\eta\to\pi^0\pi^0\gamma\gamma$ decay width for $z\simeq 1/4$,
but the possibility of observing a departure from the dominant
$\pi^0$-pole background (dashed line in Figs.1 and 2) due to chiral-loop and
vector-meson effects seems clearly open in the range $0.08 < z < 
0.18$.

As in our previous analysis on $\eta\to \pi^0\gamma\gamma$, we have now
achieved a rather complicated description for the closely related $\eta\to
\pi^0\pi^0\gamma\gamma$ amplitude, containing various contributions whose
relative weights are difficult to disentangle. To clarify this issue and to
test our results, we have computed analytically these separate contributions
at the higher end of the $\gamma\gamma$-spectrum, $z\simeq 1/4$ if one
takes $m_\eta \simeq 4 m_{\pi^0} \equiv 4 m$. Here, the two $\gamma$'s fly apart
with the same helicities and energies, $E_\gamma \simeq m$. 
Working with $\eta\to \pi^0\pi^0\gamma\gamma$ helicity amplitudes 
one thus obviously has $A^{+-}= A^{-+}=0$ and $A^{++} = A^{--}$. 
~For the latter, our four 
eqs.(\ref{pi0pole},\ref{etatail},\ref{1pietaphys},\ref{VMD}) imply the 
following four contributions displayed according to the 
decomposition in eq.(\ref{4contrib}),
\begin{eqnarray}
\label{check}
A^{\pm \pm}&\simeq& {\sqrt{2}\over 9 \sqrt{3}}{e^2 m^2 \over \pi^2f_\pi^2 
f_\eta} \Biggl\{ -(1+U+V+W){f_\eta\over f_\pi }
- 0.5 (1+ C_{loops}) - 3.9 {m^2\over f^2} R + 0.20
{f_\eta\over f_\pi} \Bigl({g\over \pi}\Bigr)^4\Biggr\} \nonumber \\
&\simeq& {\sqrt{2}\over 9 \sqrt{3}}{e^2 m^2 \over \pi^2f_\pi^2 
f_\eta}\Biggl\{-(1.51 + 0.18 i){f_\eta\over f_\pi }
-0.1 + ( 1.4 - 3.1 i) 
+ 0.81 \Biggr\},
\end{eqnarray}
where the following approximations have been made: $\Delta m_K^2 \simeq
m^2/3$, $M_V= M_{\rho,\omega}^2 \simeq 33 m^2$ and  $R\equiv R(4 m^2, -2m^2,
m^2)\simeq -0.36 + 0.84 i$. 
In this same spirit, we have worked out an approximate expression 
for the diphoton spectrum, valid for  $z\to 1/4$, i.e., 
for non-relativistic pions, 
which serves to fix the global normalization. We find
\begin{equation}
\label{diphoton}
{d \Gamma(\eta\to\pi^0\pi^0\gamma\gamma)\over d z}\bigg|_{z \simeq 1/4} \simeq
{\sqrt{2} m^3 \over (4 \pi)^4 } (|A^{++}|^2 + |A^{--}|^2) 
 {z^{1/4} (1-2 m/m_\eta^{exp} -\sqrt{z})^2 \over (1+2\sqrt{z})^{3/2}},
\end{equation}
where the experimental value of the $\eta$-mass, $m_\eta^{exp}$, has 
been kept in the final factor of the phase-space term in order
to get a reliable result at the end of the spectrum.
We can then approximately 
reproduce all our results in Figs.1 and 2 for $z\to 1/4$, as it is 
shown by the symbols (diamonds) at the end of the spectrum, corresponding to
the $\pi^0$-pole (Fig. 1) and the total (Fig. 2) contributions.

We now turn to briefly discuss the charged channel
$\eta\to\pi^+\pi^-\gamma\gamma$. The isospin conserving, 
tree level matrix element is simply found to be
\begin{eqnarray}
\label{charged}
A(\eta_8 \to   \pi^+ \pi^-\gamma\gamma)&& =
\frac{e^2}{\sqrt{6}f^3\pi^2} \epsilon_{\mu\nu\alpha\beta}
\epsilon_1^\mu k_1^\nu
\Bigl(\frac{1}{6} \frac{m^2_\pi}{s-m_\eta^2} 
\epsilon_2^\alpha
k_2^\beta \nonumber \\ 
&& + (\epsilon_2^\alpha +\frac{p_+ \cdot \epsilon_2}
{p_+ \cdot k_2} p_-^\alpha + \frac{p_- \cdot \epsilon_2}
{p_- \cdot k_2} p_+^\alpha) p_3^\beta \Bigr) 
+\Bigl[{k_1\choose \epsilon_1} \leftrightarrow {k_2\choose \epsilon_2}
\Bigr],
\end{eqnarray}
as in the analysis by Kn\"ochlein et al. \cite{KNOCH}. This amplitude 
contains two independent, gauge invariant pieces, similar to those 
entering the $\gamma\gamma\to \pi^+\pi^-\pi^0$ amplitude \cite{TABBC}.
In the latter process, these two pieces tend to cancel due to an almost 
perfect destructive interference and thus making the next $O(p^6)$ 
corrections ---which spoil the previous almost perfect cancellation--- 
numerically very important. Instead, in the tree-level $\eta_8$ decay 
amplitude (\ref{charged}), the second piece ---which corresponds 
to a brehmstrahlung process--- is singular for vanishing
photon energies, and thus dominates over the former. 
Thanks to this, in order to compute $O(p^6)$ corrections, 
one can reasonably restrict oneself only to those corresponding 
to the dominant bremsstrahlung piece, which can easily be  
obtained applying Low theorem to the
$O(p^6)$ contribution for $\eta\to \pi^+\pi^-\gamma$ \cite{BBC}. 
In addition, one also has to include the isospin violating contribution,
mediated by a $\pi^0$ pole via the decay chain $\eta \to 
\pi^+\pi^-\pi^0 \to \pi^+\pi^-\gamma\gamma$, 
which can be estimated, at $O(p^6)$ and including $\eta$-$\eta'$-
mixing effects, along the same lines 
as in the neutral channel. 

In Fig. 3 we show the di-photon mass spectrum for the 
$\eta\to \pi^+\pi^-\gamma\gamma$ decay at $O(p^6)$ 
(solid line); the  tree-level prediction for the 
$\eta_8\to \pi^+\pi^-\gamma$ process is also shown (dashed line). 
The latter is in reasonable agrement with the results in 
ref. \cite{KNOCH}. The former represents a substantial correction to 
(it is $\sim$ 3 times larger than) 
the tree-level result, as should be expected from the dynamics of 
the underlying $\eta\to \pi^+\pi^-\gamma$ transition. Indeed, from its 
analysis in ref. \cite{BBC} one immediatelly can deduce that  
$\Gamma (\eta_8\to \pi^+\pi^-\gamma)_{tree-level} \simeq 
(1/3) \Gamma (\eta\to \pi^+\pi^-\gamma)_{one-loop} $.
The nice interference pattern in Fig. 3 corresponds to the $\pi^0$ 
pole entering the isospin violating amplitude. But all these results 
refer essentially to the dynamics of the $\eta\to \pi^+\pi^-\gamma$ 
and $\eta\to \pi^+\pi^-\pi^0$ amplitudes rather than being genuine 
$\eta\to \pi^+\pi^-\gamma\gamma$ dynamical effects. The latter will be 
very hard to disentangle from this dominant background. In this sense, 
we agree with BI \cite{BI} in that a more detailed calculation 
seems unnecessary. Unsuccesful experimental attempts to detect 
$\eta\to \pi^+\pi^-\gamma\gamma$ decay events can be found in 
ref.\cite{PRICE}.

In summary, while the $\eta\to \pi^+\pi^-\gamma\gamma$ process seems 
scarcely interesting from the point of view of ChPT, the situation 
looks very different for the neutral $\eta\to \pi^0\pi^0\gamma\gamma$
decay mode. Although the partial width is predicted to be rather small, 
chiral-loop and VMD-counterterm effects could be detected and analyzed. 
This same qualitative conclusion has also been reached 
by other recent ChPT analyses.~

\vspace{1cm}
\noindent 
{\bf ACKNOWLEDGEMENTS} 
~ 
\vspace{0.5cm}

\noindent 
This work has been supported by CICYT, AEN95-815 and by the EURODAPHNE, 
HCMP, EEC contract \# CHRX-CT920026. P.T. has been supported by 
the EU TMR program under contract number ERB 4001GT952585.

\newpage 
\noindent
{\large \bf Appendix}
~
\vspace{1cm}

\noindent 
In this Appendix we quote the complete $O(p^6)$, isospin conserving matrix 
element for $\eta\to\pi^0\pi^0\gamma\gamma$.  The isospin violating
contribution (the background) proceeds through a $\pi^0$ 
pole and is written in eq.(\ref{pi0pole}) in the text. 

\begin{eqnarray}
&&A(\eta(p_3) \to \pi^0 (p_1) \pi^0 (p_2) \gamma (k_1)\gamma (k_2) 
)^{one-loop}_{1PI}=  
\frac{-e^2}{\sqrt 6 \pi^2 f_\pi^2 f_\eta} {1\over 16 \pi^2 f^2} 
\epsilon_{\mu \nu \alpha \beta} \nonumber \\
&& \Biggl( \cos\theta (4m_K^2-3p_{13}^2) R(p_{13}^2,-k_1\cdot p_{13},m_K^2)
(-\epsilon_1^{\mu} + \frac{\epsilon_1\cdot p_{13}}{k_1\cdot p_{13}}k_1^\mu)
k_2^\nu p_{2}^\alpha \epsilon_2^\beta \nonumber \\
&& + \cos\theta (4m_K^2-3p_{23}^2) R(p_{23}^2,-k_1\cdot p_{23},m_K^2)
(-\epsilon_1^{\mu} + \frac{\epsilon_1\cdot p_{23}}{k_1\cdot p_{23}}k_1^\mu)
k_2^\nu p_{1}^\alpha \epsilon_2^\beta \nonumber \\
&& - 2 (\cos\theta -\sqrt{2}\sin\theta)
 \Bigl(\frac{p_{12}^2}{2}R(p_{12}^2,-k_1\cdot p_{12},m_K^2)
-2(m_\pi^2-p_{12}^2)R(p_{12}^2,-k_1\cdot p_{12},m_\pi^2) \Bigr)\nonumber \\
&& (-\epsilon_1^{\mu} + \frac{\epsilon_1\cdot p_{12}}{k_1\cdot p_{12}}k_1^\mu)
k_2^\nu p_{3}^\alpha \epsilon_2^\beta \Biggr) 
+ \Biggl[{k_1\choose \epsilon_1} \leftrightarrow {k_2\choose \epsilon_2}
 \Biggr] \Biggr\}, 
\end{eqnarray}

\begin{eqnarray}
& &A(\eta \to \pi^0 \pi^0 \gamma \gamma )^{one-loop}_{\eta~tail} = 
\frac{e^2}{12\sqrt 6 \pi^2 f_\pi^2 f_\eta} \epsilon_{\mu \nu \alpha \beta}
\epsilon_1^\mu k_1^\nu \epsilon_2^\alpha k_2^\beta \nonumber \\
&& \ds \times \Biggl\{ \Biggl(1+ C_{loops}(p_{12},p_{13},k_1)\Biggr)
\Biggl(\frac{m_\pi^2}{	 s-m_\eta^2}
(\cos\theta -2 \sqrt{2} \sin\theta) +
\frac{m_\pi^2}{   s-m_{\eta'}^2} 2 \sqrt{2} \sin\theta\Biggr) \nonumber \\
&& + (p_{13} \leftrightarrow p_{23})
+ \Biggl[{k_1\choose \epsilon_1} \leftrightarrow {k_2\choose \epsilon_2}
 \Biggr] \Biggr\}, 
\end{eqnarray}
where $C_{loops}$ contains the $O(p^6)$ loop- and counter-term-contributions
which combine into the finite result

\begin{eqnarray}
&&C_{loops}(p_{12},p_{13},k_1)=  \frac{1}{16 \pi^2 f^2}\Biggl[4A( m_\pi^2) +
\frac{A(m_K^2)}{m_\pi^2}(\frac{5}{2}s+\frac{19}{6}
m_\pi^2-\frac{5}{2}p_{12}^2+\frac{10}{3}m_K^2-5p_{13}^2)\nonumber\\ 
&+&B(p_{12}^2,m_\pi^2,m_\pi^2)(-m_\pi^2+2p_{12}^2)+
\frac{p_{12}^2}{4 m_\pi^2}B(p_{12}^2,m_K^2,m_K^2)(-9s+4m_K^2-3m_\pi^2
+9p_{12}^2) \nonumber\\
&+& \frac{2}{9}B(p_{12}^2,m_\eta^2,m_\eta^2)(-\frac{7}{2}m_\pi^2+8
m_K^2) + \frac{4}{3}B(p_{13}^2,m_\eta^2,m_\pi^2)m_\pi^2\nonumber\\
&+& B(p_{13}^2,m_K^2,m_K^2)(2m_K^2+6 \frac{m_K^2}{m_\pi^2}s
+\frac{3}{2} \frac{p_{13}^2}{m_\pi^2}(-3s+3p_{13}^2-m_\pi^2-4m_K^2)) \nonumber\\
&+&192 \frac{\pi^2}{m_\pi^2}\Bigl[ 4 L^r_1 (-\frac{m_\pi^2}{3} 
(2 m_\pi^2 +5 p_{12}^2-8m_K^2-6s) 
+p_{12}^2 
(p_{12}^2 -\frac{4}{3}m_K^2-s))\\
&+& L^r_2 ( \frac{2}{3} m_\pi^2 (4m_\pi^2
+8 m_K^2 +4s-10p_{13}^2)+\frac{16}{3} m_K^2(
s-p_{13}^2)+4p_{13}^2(p_{13}^2-s)) \nonumber\\
&+&\frac{2}{3}L^r_3 (\frac{1}{3}m_\pi^2 (2m_\pi^2-5p_{12}^2+
16m_K^2+10(s-p_{13}^2)) + p_{12}^2(p_{12}^2-\frac{4}{3}m_K^2-s)
+\frac{8}{3}m_K^2(s-p_{13}^2) \nonumber\\
&+&2p_{13}^2(p_{13}^2-s)) 
+\frac{2}{3}L^r_4(m_\pi^2(7m_\pi^2+4p_{12}^2-22m_K^2-6s)+8m_K^2p_{12}^2)
\nonumber\\
&+& \frac{4}{9} L^r_5 m_\pi^2(m_\pi^2-4m_K^2-\frac{3}{2}s)
+\frac{16}{3} L^r_6 m_\pi^2(-m_\pi^2+4m_K^2) \nonumber\\
&+& \frac{64}{3} L^r_7 m_\pi^2(m_\pi^2-m_K^2)+8L^r_8 m_\pi^4
\Bigr]\Biggr],
\nonumber
\end{eqnarray}
where 
$p_{12}=p_1 + p_2$, $p_{13}=p_1 - p_3$, $p_{23}=p_2 - p_3$ and $s=(k_1+k_2)^2$.
The functions $A$, $B$ and $R$ are defined as
\begin{equation}
A(m^2)= -m^2 \ln{m^2\over \mu^2}
\end{equation}

\begin{equation}
B(p^2,m_1^2,m_2^2)= \Bigg\{ 1 -\frac{1}{2}\log
\frac{m_1^2 m_2^2}{\mu^4}+ \frac{m_2^2 -m_1^2}{2 p^2}\log\frac{m_1^2}{m_2^2}
-\frac{1}{p^2} {u_+} {u_-} \log\frac{{u_+} + {u_-}}{{u_+} - {u_-}}\Bigg\}\ ,
\end{equation}
with $u_\pm = \sqrt{p^2 - (m_1 \pm m_2)^2}$,
which, for equal masses, simplifies to

\begin{equation} 
B(p^2,m^2,m^2)=1+\beta\ln{\beta -1 \over \beta +1} -\ln{m^2\over \mu^2}, \qquad
\qquad \beta=\sqrt{1 -{4 m^2\over p^2}}
\end{equation}
and
\begin{eqnarray}
R(p^2,k\cdot p,m^2)&=&{1\over 2}+\left({1\over 2}-{p^2\over 4 k\cdot p}\right)
\left[\beta' \ln{\beta'-1\over \beta' +1}
-\beta \ln {\beta-1\over \beta +1}\right] \nonumber\\
&+& {m^2\over 4 k\cdot p}
\left[\ln^2{\beta-1\over \beta +1}-\ln^2{\beta'-1\over \beta' +1}\right],
\end{eqnarray}
with $\beta'=\ds \sqrt{1 -{4 m^2\over p^2-2 k\cdot p}}$ and $k^2=0$.

~ 

One can check that the scale $\mu$ appearing in $A$ and $B$ cancels in
$C_{loops}$, as it corresponds to a physical result. The numerical
estimation of $C_{loops}$ has been done using the central value of the
$L^r_i$ counter-terms listed in \cite{BEG}.

\listoffigures
\newpage

\newpage
\begin{figure}
\epsfig{file=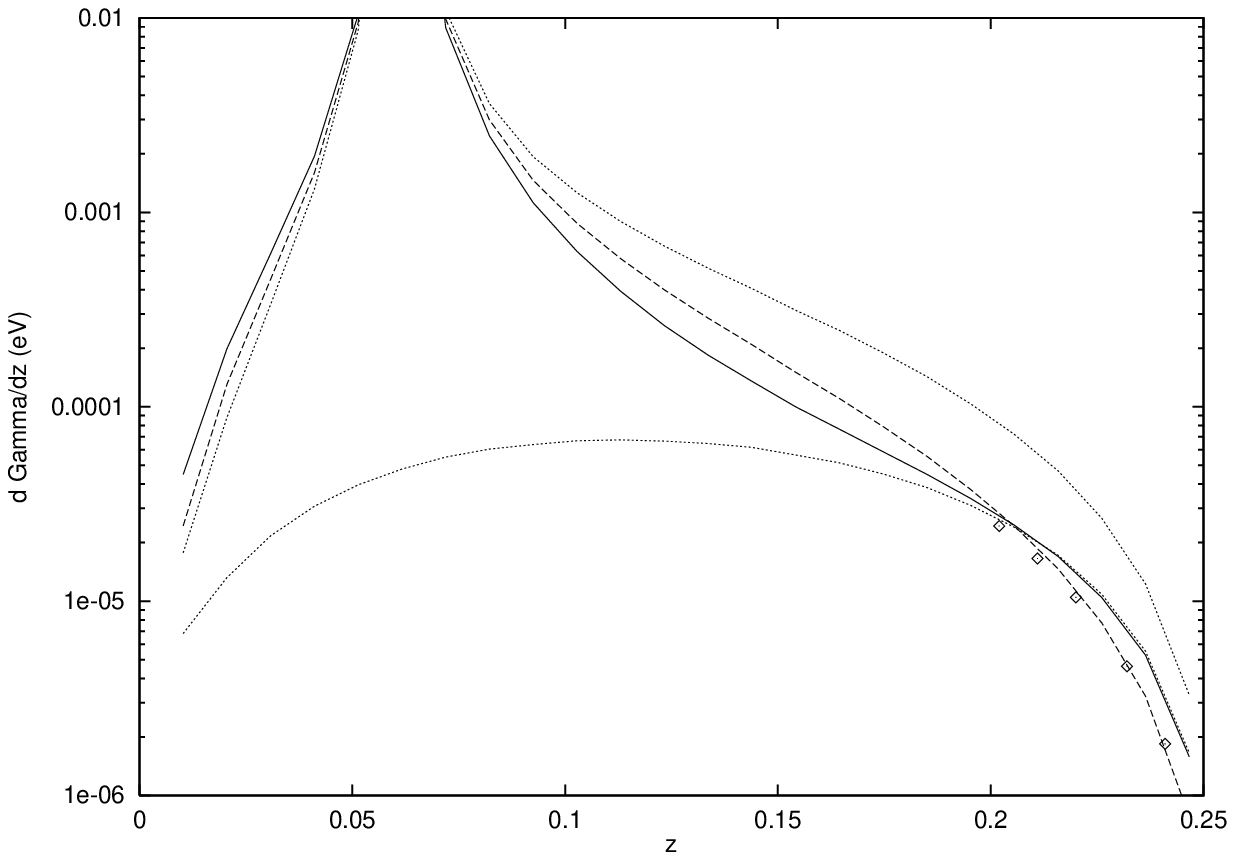,width=16cm,height=18cm,angle=0}
\caption[Diphoton mass spectrum for the $\eta\to\pi^0\pi^0\gamma\gamma$ 
decay. Partial results.] 
{Diphoton mass spectrum for the $\eta\to\pi^0\pi^0\gamma\gamma$ decay.
The dashed line corresponds to the $\pi^0$-pole contribution, eq.(5). 
The symbols (diamonds) at the end of the spectrum show the 
$\pi^0$-pole in the approximation of eq.(11) and using the first term 
of eq.(10).
The lowest dotted line is the 1PI result, eq.(7), for $\eta_8$. The solid
line is our result when adding  eqs.(5) and (7).  
The upper dotted line (denoted by BI in the text)
shows the result of subtracting eqs.(5) and (7).} 
\end{figure}
\begin{figure}
\epsfig{file=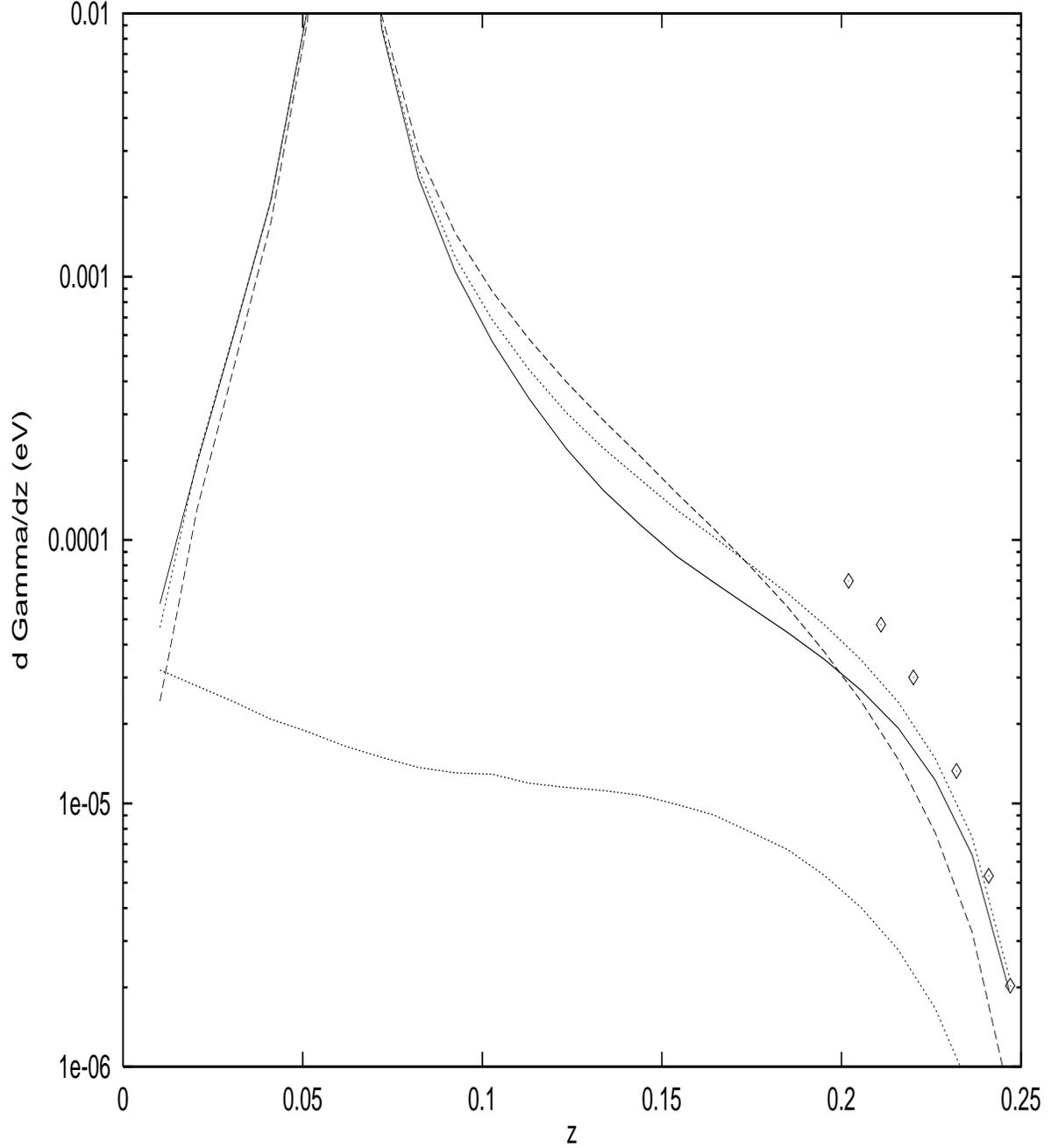,width=16cm,height=18cm,angle=0}
\caption[Diphoton mass spectrum for the $\eta\to\pi^0\pi^0\gamma\gamma$ 
decay. Final results.]
{Diphoton mass spectrum for the $\eta\to\pi^0\pi^0\gamma\gamma$ decay.
The dashed line is  the $\pi^0$-pole contribution, eq.(5). The upper dotted
line corresponds to the $O(p^6)$ result. The lower dotted line is the VMD
contribution, eq.(9). 
The solid line corresponds to the total contribution, listed in eqs.(4,5,6,8
and 9).  The symbols (diamonds) at the end of the spectrum show the
total result using the approximation of eqs.(10) and (11).}
\end{figure}
\begin{figure}
\epsfig{file=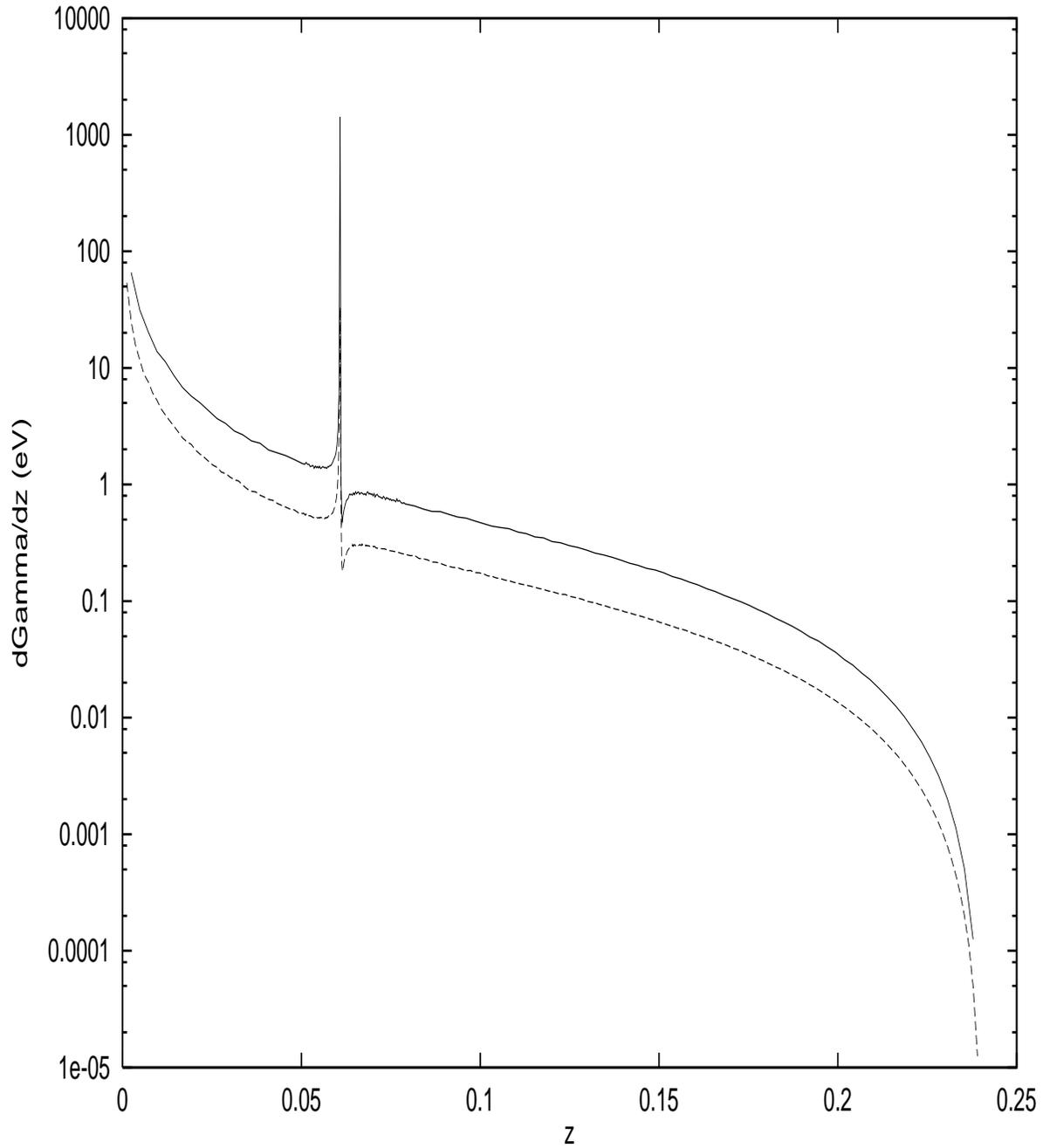,width=16cm,height=18cm,angle=0}
\caption[Diphoton mass spectrum for the $\eta\to\pi^+\pi^-\gamma\gamma$
decay.]
{Diphoton mass spectrum for the $\eta\to\pi^+\pi^-\gamma\gamma$
decay. The solid line is the $O(p^6)$ result. The dashed
curve corresponds to the tree-level result for $\eta_8$, eq.(12).}
\end{figure}
\end{document}